\documentclass[journal=jacsat,manuscript=article]{achemso}

\usepackage[version=3]{mhchem} % Formula subscripts using \ce{}

\usepackage{latexsym}
\usepackage{color}
\usepackage{achemso}
\usepackage{amssymb}
\usepackage{pifont}
\newcommand{\cmark}{\ding{51}}%
\newcommand{\xmark}{\ding{55}}%

\author{Elias Torres Alonso$^{**}$}
\author{George Karkera$^{**}$}
\author{Gareth F. Jones}
\author{Monica F. Craciun}
\author{Saverio Russo}
\email{s.russo@exeter.ac.uk}
\affiliation[CfGS]
{Centre for Graphene Science, College of Engineering, Mathematics and Physical Sciences, University of Exeter, Exeter EX4 4QF, United Kingdom}

\title[An \textsf{achemso} demo]
  {Homogeneously bright, flexible and foldable lighting devices with functionalised graphene electrodes%\footnote{A footnote for the title}
  }

\begin{document}



\begin{abstract}
Alternating current electroluminescent technology allows the fabrication of large area, flat and flexible lights. Presently the maximum size of a continuous panel is limited by the high resistivity of available transparent electrode materials causing a visible gradient of brightness. Here, we demonstrate that the use of the best known transparent conductor FeCl$_{3}$-intercalated few-layer graphene boosts the brightness of electroluminescent devices by 49$\%$ compared to pristine graphene. Intensity gradients observed for high aspect ratio devices are undetectable when using these highly conductive electrodes. Flat lights on polymer substrates are found to be resilient to repeated and flexural strains.
\end{abstract}


\newpage

Displays and screens are some of the most important devices within the field of optoelectronics. Large-area displays are ubiquitous in advertising, ambient lighting, television, computer screens and almost every device with an interface for its control. Current research is focused on making these devices suitable for the next generation of flexible and foldable optoelectronics. In particular, foldable displays and lighting systems would allow expandable screens for smart phones and electronic paper, wearable optoelectronics, rollable or collapsible wallpaper lights, and biocompatible light sources for medical devices. To date, alternating current electroluminescent (ACEL) technology uniquely enables the realization of flat, large-area ($\approx$ 1m$^2$) and flexible light sources\cite{Bredol2010}. ACEL devices have good contrast and uniform brightness. They can display images with high resolution when employing nano-patterned electrodes and they can withstand mechanical shocks as well as a wide range of temperatures\cite{Wang2011,Allieri2002}. Presently, the maximum size of ACEL lighting panels is still limited to less than a square meter under practical operational conditions. This is due to the visible gradient of brightness which develops across a panel induced by the high sheet resistance (typically larger than 100$\Omega/\Box$) of the transparent electrode materials\cite{Chukova1967,Craciun2015}. Highly resistive electrodes also require a high operational voltage and frequency, leading to increased power consumption and accelerated device degradation\cite{Bredol2010,Craciun2015}. 

Poly(3,4-ethylenedioxythiophene) polystyrene sulfonate (PEDOT:PSS) is a transparent and conductive polymer widely used by the printed electronics industry owing to its high mechanical flexibility. However, PEDOT:PSS has a high sheet resistance of 850$\Omega/\Box$ for an optical transparency of 90\%\cite{Nagata2011,Lang2009,Hecht2011}, an unwanted blue tinge\cite{Hecht2011} and limited environmental stability\cite{Nagata2011,Lang2009,Hecht2011}. When exposed to relative humidity of just 40\%, the uptake of water by the hygroscopic PSS leads to the development of shear lips when the films are deformed, greatly increasing the sheet resistance\cite{Lang2009}. Moreover, the absorption of water combined with the acidity of PSS is a well-known cause of degradation in device components\cite{Jorgensen2008,rand2014organic}. Atomically thin conductors such as graphene possess many attractive properties for transparent electrodes, which overcome some of the limitations of PEDOT:PSS. Single layer graphene only absorbs $\approx$ 2.3\% of visible light and it is highly flexible\cite{Craciun2015}. However, the sheet resistance of this single layer of carbon atoms exceeds 1k$\Omega/\Box$ in its pristine form\cite{Craciun2015}, making it unsuitable for large-area ACEL devices. Meshes of metallic nanowires are an emergent technology characterized by lower resistivity than PEDOT:PSS and graphene. However, these nanowires have several problems: the nanowire dimensions have to be accurately control to avoid failure due to mechanical stresses\cite{Wu2012}; they suffer from optical haze due to different size and spacing from this nanowires \cite{Hecht2011}; lack electrical stability due to current-induced\cite{Gold} and electromigration-induced structural damaged \cite{JZhao}. Carbon nanotubes (CNT) meshes have been employed in ACEL devices\cite{Schrage2009}, but their high resistivity (5k$\Omega/\Box$) prevents their use in large-area devices. The search for a low resistance, transparent, flexible and stable material is still ongoing. This long-sought material would take lighting technology from the rigid confines of a glass bulb or an electronic semiconductor chip to emerging flexible substrates such as textile fibres\cite{Neves2015} and paper technologies.

In this work we present the use of the outstanding transparent conductor FeCl$_3$ intercalated few-layer graphene (FeCl$_3$-FLG)\cite{Craciun2015,Khrapach2012,Bointon2014} for highly uniform, flexible and foldable ACEL devices. FeCl$_3$-FLG has a sheet resistance as low as 20$\Omega/\Box$ and an optical transparency of 77\%  when using CVD-grown graphene\cite{Bointon2014}. It is highly stable against 100\% humidity and temperature up to 150\textsuperscript{o}C \cite{Wehenkel2015}, and was shown to be a promising material for work function matched transparent electrodes in photovoltaic and organic light emitting devices\cite{Bointon2014,Craciun2015} and photodetectors \cite{Withers2013}. We compare the performance of this emerging material against the widely-used conductive polymer PEDOT:PSS and other graphene electrodes such as single-layer graphene (SLG) and FLG. Our experimental findings demonstrate that FeCl$_3$-FLG-based ACEL devices display up to 49\% higher electroluminescent intensity at low operational voltage (from 40V to 180V) compared to devices embedding pristine FLG. Contrary to ACEL devices employing alternative electrode materials, the use of FeCl$_3$-FLG transparent electrodes does not show any measurable gradient of the intensity of the emitted light in flat panels 4cm long and 0.8 mm wide, owing to the low sheet resistance of FeCl$_3$-FLG\cite{Bointon2014}. This finding is in stark contrast to the experimentally measured 70\% reduction in the intensity of the emitted light measured in ACEL devices embedding FLG electrodes. Finally, we demonstrate the performance of this outstanding transparent electrode in flexible and foldable lighting devices fabricated on polyethylene terephthalate (PET). These flexible ACEL devices are both highly flexible, durable and foldable, showing no degradation in performance under flexural strains beyond the common requirements of wearable electronics applications and over more than 1000 bending cycles. Our results pave the way for the next generation of energy efficient, flexible, high performance and large-area optoelectronics based on highly transparent and conductive electrodes.

Figure 1a shows a schematic of the inner structure of the studied ACEL devices. State-of-the-art graphene-based transparent electrode materials are employed and their performance compared to that of commercial PEDOT:PSS. These materials are (1) high quality pristine single- and (2) few-layer graphene grown by chemical vapour deposition and (3) FeCl$_3$-intercalated few-layer graphene. All graphene materials were transferred from Cu or Ni to glass microscope slides via a standard PMMA-assisted wet etching technique (see Experimental Section). Intercalation of FLG with FeCl$_3$ was carried out in vacuum (2$\times 10^{−4}$\,mbar) positioning the anhydrous FeCl$_3$ powder and the few-layer graphene substrate in different temperature zones inside a glass tube. The FLG and the
powder are heated for 7.5 hours at $360\,^\circ$C and $310\,^\circ$C, respectively, with a rate of $10\,^{\circ}$C/min
during the warming and cooling of the two zones. The intercalation of FeCl$_3$ and the doping of graphene are characterized analysing the up shift of the G-peak in the Raman spectra of graphene (see supplementary information). Layers of light-emitting phosphor material (doped zinc sulphide), dielectric (barium titanate, BaTiO$_3$) and bottom electrode (silver paste) were deposited onto the transparent electrode by spin coating and dried in air at 120\textsuperscript{o}C on a hot plate. The top and bottom electrodes were contacted as shown in Figure 1a. In order to accurately examine the decay in emitted light intensity without the use of impractically large display panels, we designed electrodes with two different aspect ratios (length:width of 4:1 and 30:1) and direct electrical contact at only one end of the transparent conductor (i.e. with no bus-bar around the perimeter). The thicknesses of the phosphor and dielectric layers were calibrated to the supplier specifications of 30$\mu$m and 25$\mu$m respectively (see cross section image in Figure 1a). Upon applying an AC voltage (681 Hz and $V_{AC}\leq180V$) the flat panels emit a bright light, as shown in Figure 1b. In total, we have characterized at least 4 different ACEL devices for each transparent electrode material.

The average emission spectra in ACEL devices with FeCl$_3$-FLG and pristine FLG for $V_{AC}=180V$ are reported in Figure 2a. It is apparent that the devices with FeCl$_3$-FLG electrodes display a significantly enhanced brightness compared to the case of pristine FLG, reaching up to a 49\% increase at the spectral peak ($\approx$ 500nm). The emission spectra and total emitted light intensity of ACEL devices with FeCl$_3$-FLG is also found to be stable in time, with no detectable change over a period of 4 months when storing the device in ambient conditions (see supplementary information). In both cases, the emission spectra have a very similar shape stemming from the flat optical transmittance of FeCl$_3$-FLG and pristine FLG in the emission range of the phosphor layer\cite{Bointon2014,Khrapach2012}. Hence, we fit the spectra of each ACEL device with the superposition of the four known Gaussian emission bands ($\lambda=$438nm, 464nm, 497 and 525nm) corresponding to the intraband transitions originating from sulphur vacancy states and a variety of surface states \cite{Manzoor2003}. Broadening of these peaks is mainly determined by variations in the size \cite{Ibanez2007} and crystalline phase \cite{Allieri2002} of phosphor particles. The average intensity for FeCl$_3$-FLG electrodes was found to be systematically higher (see Supporting Information) than that of pristine FLG electrodes at any given wavelength and for any applied voltage in the range $40V<V_{AC}<180V$, see Figure 2b. This is to be expected since the sheet resistance of FeCl$_3$-FLG is approximately ten times lower than that of FLG, leading to a higher average voltage across the electrode.

In Figure 2c the light intensity at the emission peak ($\approx$ 500nm) for devices embedding FLG and FeCl$_3$-FLG electrodes is plotted as a function of $V_{AC}$. The bias dependence of the average light intensity is well described by the relationship between ACEL brightness ($B$) and voltage ($V$)\cite{Bredol2010}: $B=B_{0}\exp\left[-(V_{0}/\sqrt{V})\right]$, where $V_0$ and $B_0$ are empirical constants. The excellent agreement between experimental data and the fit to this Alfrey-Taylor relation demonstrates that the light emission is governed by a Mott-Schottky-type barrier, as previously reported \cite{ALFREY1955,Bredol2010,Kouyate1992}. The pronounced gain in the average brightness measured in ACEL flat panels embedding FeCl$_3$-FLG transparent electrodes is also apparent to the naked eye when biasing the devices with the same $V_{AC}$, see Figure 2d.

To directly test the brightness gradient introduced by the different transparent electrode materials, we fabricated devices with high aspect ratios (length:width, 30:1) with FeCl$_3$-FLG, pristine FLG, SLG and PEDOT:PSS transparent electrodes. In this geometry the non-uniform intensity of the emitted light brought about by the voltage drop along the transparent electrodes is readily accessible. Figure 3 shows the normalized intensity of the emitted light collected at spots of 5mm diameter along the length of the device. This brightness was normalised to that of the point closest to where the transparent electrode is contacted. These measurements show that all of the transparent electrode materials, with the exception of FeCl$_3$-FLG, display a sizeable drop of emitted light intensity moving away from the point where the external $V_{AC}$ is applied. Single layer graphene electrodes display the most dramatic drop in intensity, that is up to $\approx$ 70\% reduction at 27mm from the contact. In contrast, FeCl$_3$-FLG does not display any measurable intensity gradient under the same experimental conditions. We have confirmed this surprising finding in all 11 studied ACEL devices (see Supporting Information).

The observed spatial dependence of the light intensity for the different transparent electrode materials correlates well with their intrinsic resistivity. Indeed, the most resistive transparent electrodes studied in this work are single layer graphene  ($\approx$ 1k$\Omega$/$\Box$) and this material gives the largest light intensity gradient in ACEL devices. Electrodes of FeCl$_3$-FLG with just $\approx$ 20$\Omega$/$\Box$ produce no measurable light intensity gradient since these electrodes exhibit a negligible voltage drop over the studied length scales ultimately leading to a uniform light intensity across the whole surface of the devices.

Lastly, to demonstrate the capability of FeCl$_3$-FLG as a flexible and foldable electrode, we study the performance of FeCl$_3$-FLG based ACEL devices fabricated on a plastic substrate when subjected to a range of bending and strain conditions. Figure 4a shows a sequence of pictures taken under the same conditions of exposure time and diaphragm for a representative device undergoing a folding cycle with a radius down to less than 3mm and a light emitting area of $\approx$ 4cm$^2$ (see supplementary video information). It is apparent to the eye that the intensity of the emitted light does not change during the entire folding cycle. This is quantitatively presented in the plot of Figure 4b, where we show the measured intensity of the emitted light to be independent of the flexural strain in the FeCl$_3$-FLG electrode. The strain is defined as $\epsilon=(d-2t_s)/(2r)$, where $d$ is the total thickness of the device, $t_s$ is the thickness of the substrate and $r$ is the radius of curvature, see supplementary information. The same graph also shows the corresponding values of radius of curvature on the upper x-axis. In our experiment, the curvature radius is purposefully varied in the range $10<r<35$mm since an electronic device wrapped as a wristband would have $r\approx 25$mm (based on an average adult thickness).

The light intensity was also found to be unchanged after performing more than 1000 bending cycles of the device around a tube with 7mm radius, as shown in the inset of Figure 4c. For comparison, the performance of a PEDOT:PSS-based device under the same conditions is also shown. This demonstrates that FeCl$_3$-FLG is as flexible and durable to mechanical cycles as PEDOT:PSS, known to be a highly flexible and resilient conductor. However, PEDOT:PSS lacks environmental stability whereas previous experiments have shown that FeCl$_3$-FLG can withstand high relative humidity of 100\% at room temperature for at least 25 days\cite{Wehenkel2015}. Unlike PEDOT:PSS, FeCl$_3$-FLG has flat optical absorption and there are no known compatibility issues between FeCl$_3$-FLG and other commonly used optoelectronic materials. A summary of properties of emerging transparent electrode materials and state-of-the-art systems used by the industry is presented in Table 1 which demonstrates the outstanding features of FeCl$_3$-FLG.

In summary, we have conducted a systematic study of the light emission in electroluminescent devices with a range of graphene-based transparent electrodes and compared their performance to that of PEDOT:PSS which is one of the most widely used conductive polymers in printed electronics. Our experiments demonstrate that the implementation of FeCl$_3$-FLG transparent electrodes in ACEL devices results in a dramatic enhancement of the intensity of the emitted light up to 49\% at low operational voltage (from 40V to 180V) compared to devices embedding pristine few-layer graphene. We show that owing to the intrinsic low electrical resistivity characterizing FeCl$_3$-FLG, there is no detectable reduction of the emitted light intensity moving away from the contacts. This is in stark contrast to all the other studied materials, which display a sizable light intensity reduction - up to a 70\% decrease for single layer graphene electrodes 4\,cm long and 0.8\,cm wide. Finally, we demonstrate that ACEL devices embedding FeCl$_3$-FLG electrodes are mechanically flexible and most importantly foldable. No change in the intensity of the emitted light is detected even after more than 1000 bending cycles with bending radius of just 7mm. These findings demonstrate that FeCl$_3$-FLG is highly suitable for large-area, flexible and foldable lighting technologies, paving the way to conceptually new wearable optoelectronic applications.

\begin{suppinfo}

\begin{itemize}
  \item Transparent electrodes and light emitting layer; spectroscopy and light emission characterization; Raman characterization; electroluminescent emission spectra; strain set-up and strain calculation, device stability over time (PDF). 
  \item Video of a FeCl$_{3}$-FLG EL device being folded (AVI).
\end{itemize}

\end{suppinfo}

\section{Author information}
\textbf{Corresponding Author}
s.russo@exeter.ac.uk

\textbf{Author Contributions}
$^{**}$ E.T.A. and G.K. contributed equally to this work

\textbf{Notes} The authors declare no competing financial interest.

\begin{acknowledgement}
The authors thank D.J. Wehenkel for suggestions on the specific electroluminescent materials and A. De Sanctis for help with optical characterisation, J. Bannerman for assistance with SEM images, P. Wilkins and A. Woodgate for workshop assistance. S. Russo and M.F. Craciun acknoweldge financial support from EPSRC (Grant no. EP/J000396/1, EP/K017160/1, EP/K010050/1, EPG036101/1, EP/M001024/1, EPM002438/1) and from the Leverhulme Trust (Research grant title Quantum Drums).
\end{acknowledgement}

\newpage

\begin{figure}
\begin{center}
\includegraphics[width=\textwidth]{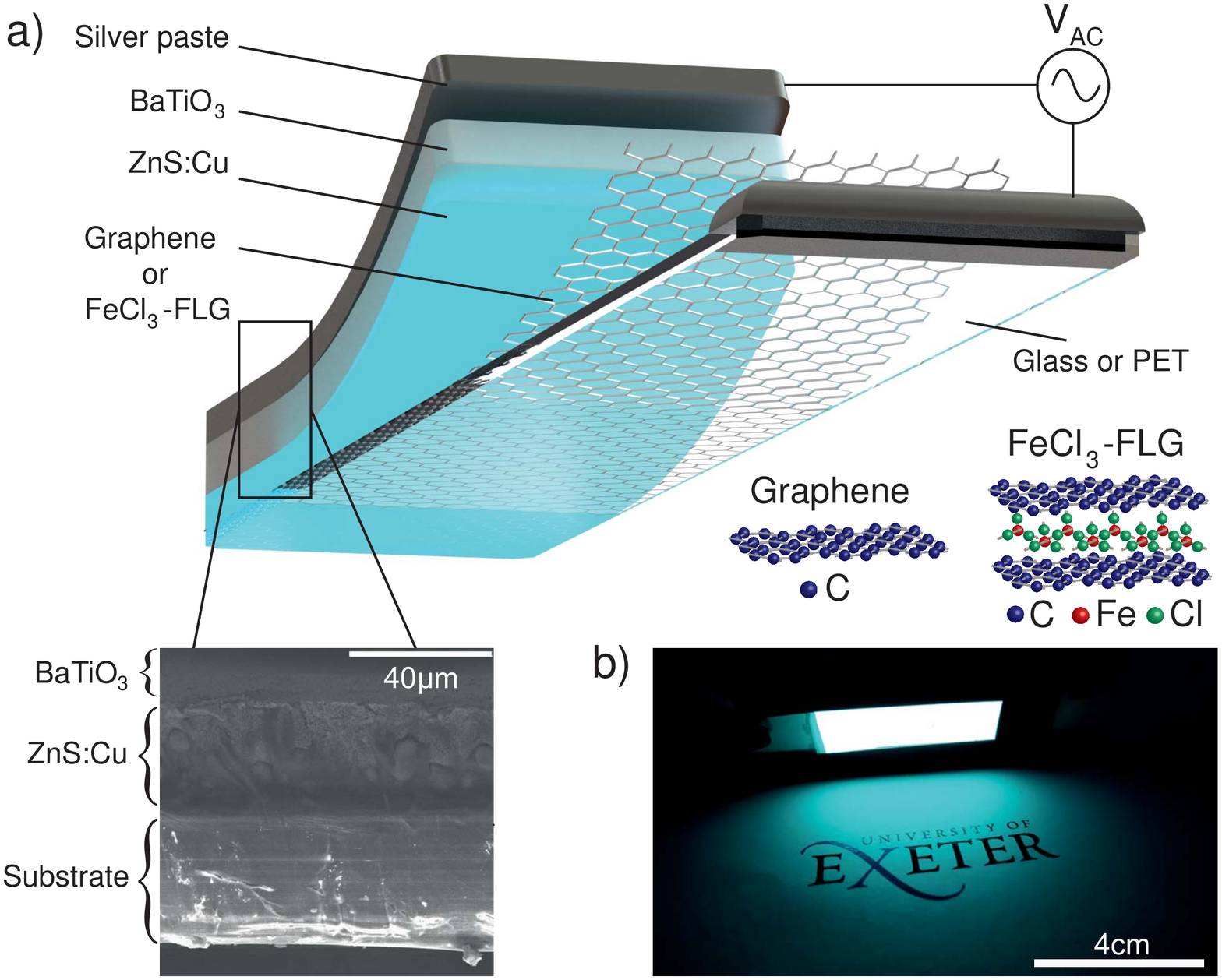}
\caption{a) shows a schematic representation of the layered structure of an ACEL device with a zoom on the cross section showing a scanning electron microscope image of the actual sequence of materials. The atomic structure of the transparent electrodes based on graphene and FeCl$_3$-intercalated few-layer graphene is shown in the cartoon reported in the inset. b) shows a picture of an ACEL device with FeCl$_3$-FLG transparent electrode operating in ambient condition with an applied voltage bias $V_{AC}=180$V and frequency 681Hz.}
\label{Fig1}
\end{center}
\end{figure}

\newpage

\begin{figure}
\begin{center}
\includegraphics[width=\textwidth]{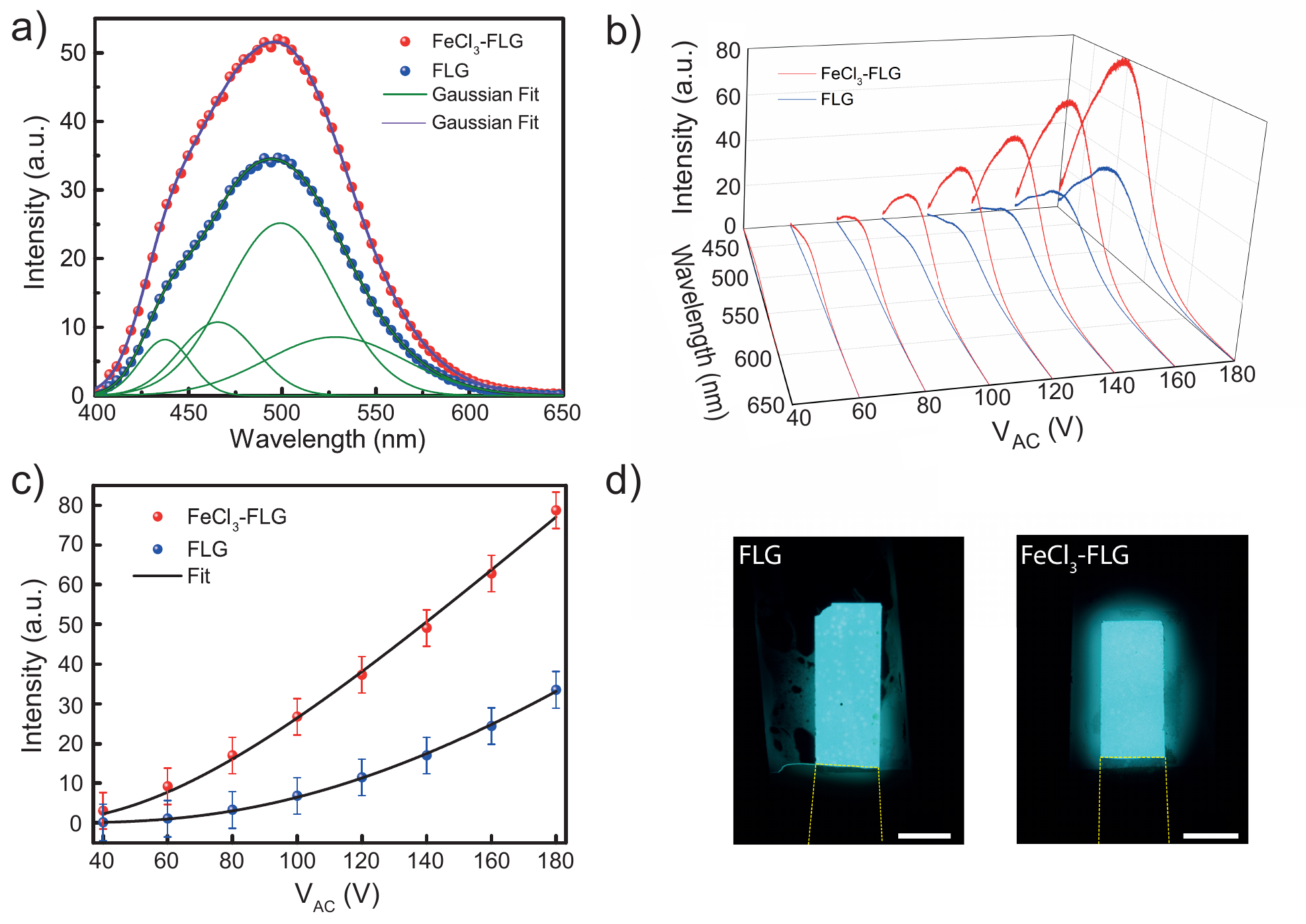}
\caption{a) Emitted light intensity of ACEL devices embedding FeCl$_{3}$-FLG and FLG electrodes. The 4 Gaussian curves (green) correspond to the fit of the FLG-based ACEL spectrum as discussed in the manuscript. The cumulative fits for the FeCl$_3$-FLG- and FLG-based device are shown in black and red respectively. b) Emitted light intensity of FeCl$_{3}$-FLG- and FLG-based devices for different applied bias voltages. c) Shows a graph of the peak intensity ($\approx$ 500nm) of the spectra shown in panel (b) as a function of applied bias voltage. The squares are experimental data and the continuous line is a fit to the Alfrey-Taylor relation. d) Picture of the FeCl$_{3}$-FLG- and FLG-based devices lighting up with V$_{AC}$=100V. The white scale bar corresponds to 1 cm.}
\label{Fig2}
\end{center}
\end{figure}

\newpage

\begin{figure}
\begin{center}
\includegraphics[width=9cm]{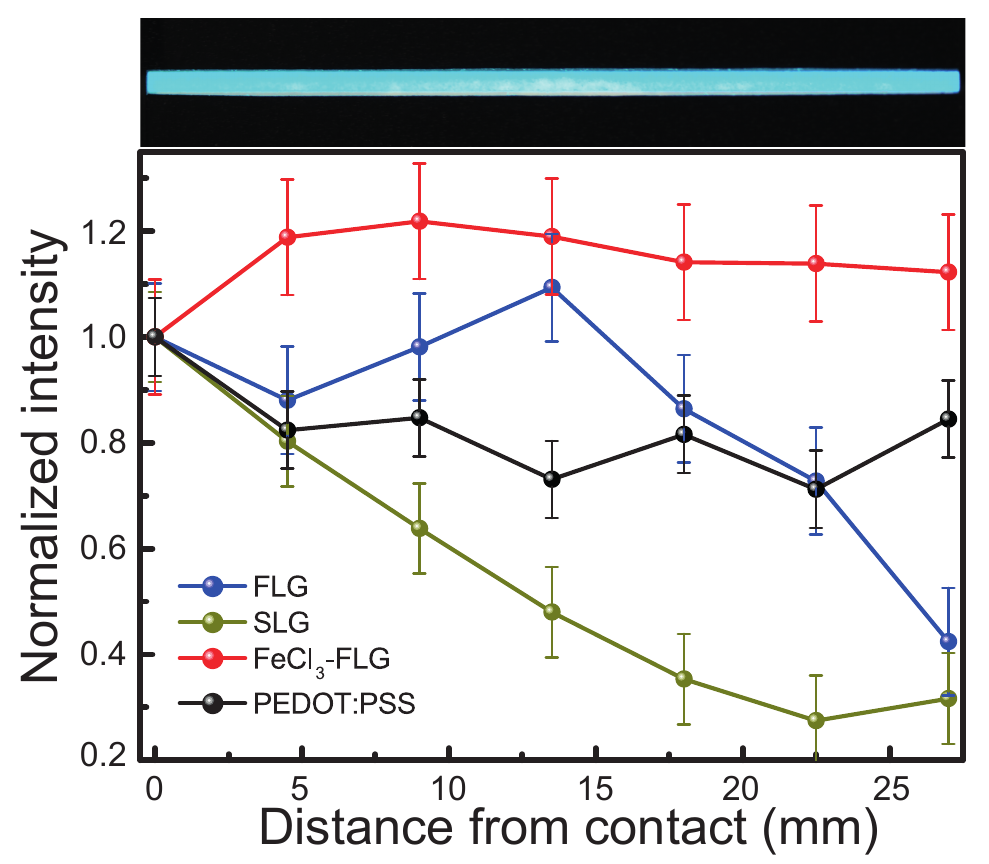}
\caption{Normalized intensity as function of distance from the contact for high aspect ratio devices as shown in the picture. The points at which the light intensity was measured are mapped to the picture of an FeCl$_3$-FLG-based device above.}
\label{Fig3}
\end{center}
\end{figure}

\newpage

\begin{figure}
\begin{center}
\includegraphics[width=\textwidth]{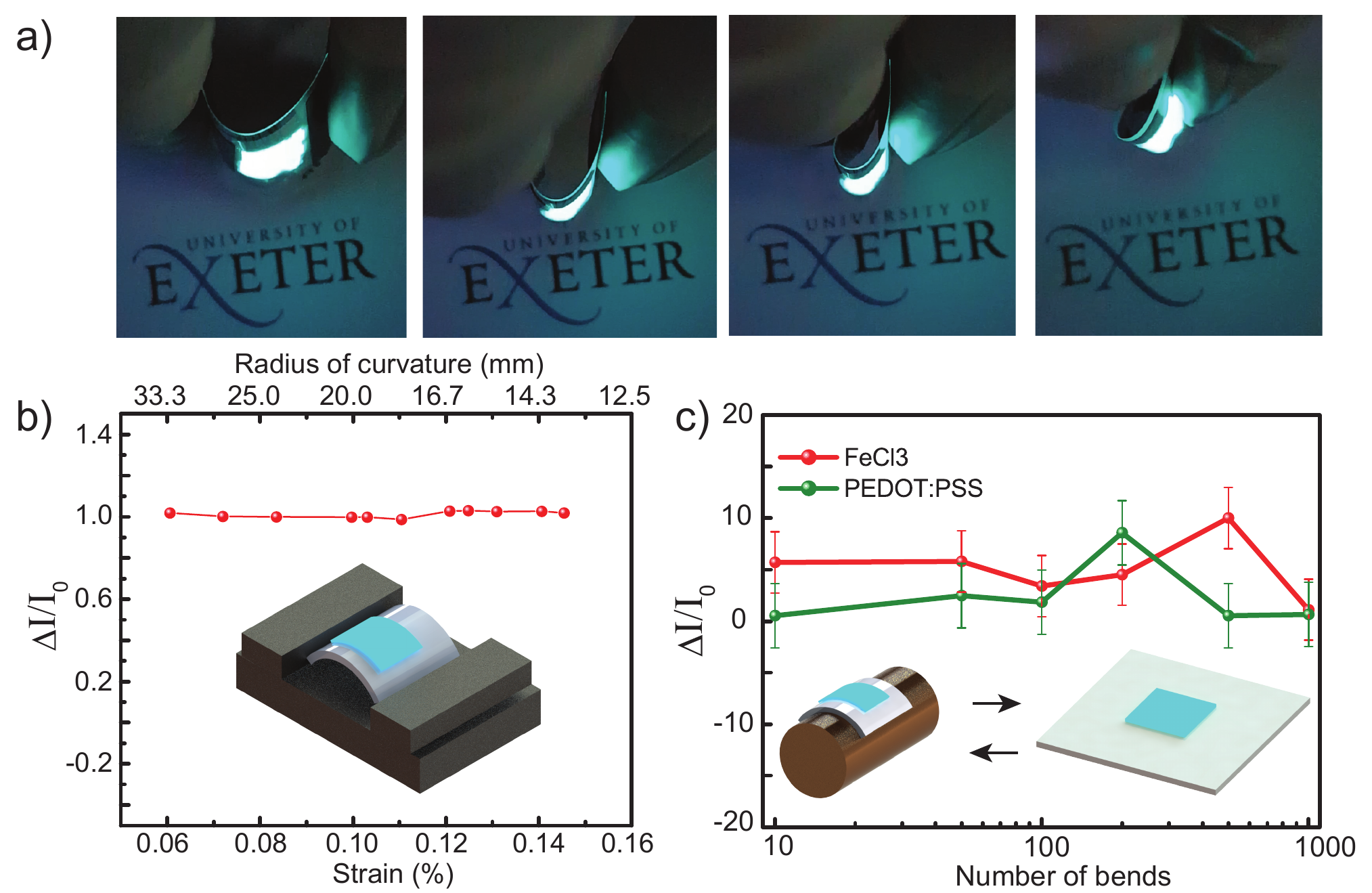}
\caption{a) Images showing an ACEL device with FeCl$_3$-FLG transparent electrode fabricated on PET substrate undergoing increasing bending. (b) Normalized emitted light intensity as function of the radius of curvature and strain applied to the FeCl$_3$-FLG electrode. The inset shows a schematic of the set-up used to apply the strain to the ACEL device. c) Normalized emitted light intensity as function of number of bending cycles for the FeCl$_3$-FLG and PEDOT:PSS transparent electrodes embedded in ACEL devices. The inset shows a schematic of the bending/measurement procedure.}
\label{Fig4}
\end{center}
\end{figure}

\newpage

\begin{table}
  \caption{Comparison of transparent electrode materials for ACEL devices}
  \label{tbl:example}
  \begin{tabular}{lllll}
    \hline
    Electrode & R$_{\Box}$ & Tr (550nm) & Flexibility & Ambient Stability\\
    \hline
    FeCl$_3$-(Ni-CVD)FLG\cite{Khrapach2012}   & {\color{green}20$\Omega/\Box$}  & {\color{green}77\%} & {\color{green}\cmark} & {\color{green}\cmark}  \\
    ITO\cite{De2010} & {\color{green}10$\Omega/\Box$} & {\color{green}85\%} & {\color{red}\xmark} & {\color{green}\cmark} \\
    SLG\cite{Craciun2015}  & {\color{red}1000$\Omega/\Box$} & {\color{green}97\%} & {\color{green}\cmark} & {\color{green}\cmark}\\
    FLG (Ni-CVD)\cite{Kim2009} & {\color{red}280$\Omega/\Box$} & {\color{green}80\%} & {\color{green}\cmark} & {\color{green}\cmark} \\
    PEDOT:PSS\cite{Nagata2011,Lang2009,Hecht2011} & {\color{red}850$\Omega/\Box$} & {\color{green}91\%} & {\color{green}\cmark} & {\color{red}\xmark}\\
    CNT\cite{Schrage2009} & {\color{red}5800$\Omega/\Box$} & {\color{green}75\%} & {\color{green}\cmark} & {\color{green}\cmark} \\
    Metallic nanowires\cite{Hu2010} & {\color{green}20$\Omega/\Box$} & {\color{green}80\%} & {\color{green}\cmark} & {\color{red}\xmark} \\
    \hline
  \end{tabular}
\end{table}

\end{document}